\documentclass[twocolumn]{aastex61} 
\usepackage[utf8]{inputenc}

\usepackage{lineno}
\usepackage{natbib}
\usepackage{graphicx}
\usepackage{amsmath} 
\usepackage{mathtools}

\shorttitle{Ultra-Short-Period Planets and FU Ori Outbursts}
\shortauthors{Becker et al.}

\begin{document}

\title{Migrating Planets into Ultra-Short-Period Orbits during Episodic Accretion Events} 

\correspondingauthor{Juliette Becker}
\email{jbecker@caltech.edu}

\author[0000-0002-7733-4522]{Juliette C. Becker}
\altaffiliation{51 Pegasi b Fellow}
\affiliation{Division of Geological and Planetary Sciences, Caltech, Pasadena, CA, 91125}
\author[0000-0002-7094-7908]{Konstantin Batygin}
\affiliation{Division of Geological and Planetary Sciences, Caltech, Pasadena, CA, 91125}
\author[0000-0002-8167-1767]{Fred~C.~Adams}
\affiliation{Department of Physics, University of Michigan, Ann Arbor, MI 48109}
\affiliation{Department of Astronomy, University of Michigan, Ann Arbor, MI 48109}

\begin{abstract}
Ultra-short-period (USP) planets reside inside the expected truncation radius for typical T Tauri disks. As a result, their current orbital locations require an explanation beyond standard disk migration or in situ formation. Modern theories of planet-disk interactions indicate that once a planet migrates close to the disk's inner truncation radius, Type I torques vanish or switch direction, depending on the stellar and disk conditions, so that the planet is expected to stop its orbital decay and become trapped. 
In this work, we show that that magnetically-driven sub-Keplerian gas flow in the inner disk can naturally counteract these effects and produce systems with USP planets at their observed orbital radii. The sub-Keplerian gas flow provides a headwind to small planets, and the resulting torque can overcome the effects of outward Type I migration near the co-rotation radius. For suitable disk and planet parameters, the torques due to the sub-Keplerian gas flow lead to inward migration on a rapid timescale. Over the time span of an FU Ori outburst, which moves the disk truncation radius inward, the rapid headwind migration can place planets in USP orbits. The combination of headwind migration and FU Ori outbursts thus provides a plausible mechanism to move small planets from $a=0.05-0.1$ AU down to $a=0.01-0.02$ AU. This effect is amplified for low-mass planets, consistent with existing observations.
\end{abstract}

\section{Introduction}
\label{sec:intro}
Ultra-short period (USP) planets represent a formidable challenge to current theories of planet formation and migration. These extreme objects reside interior to typical values of the disk truncation radius, where the absence of disk material renders both the formation and movement of these objects difficult. Generally defined as planets with orbital periods less than 1 day, USP planets are not common, orbiting only $\sim$0.5\% of stars \citep{SanchisOjeda2014}. Moreover, USP planets reside at the edge of the feasible parameter space for planetary orbits, with some USP planets narrowly avoiding tidal disintegration by their host stars \citep{Price2020}. 
Beyond their orbital demarcations, USP planets tend to be small, making their detection more difficult compared to those of the larger planets, especially the easily detectable hot Jupiter class of planets. Specifically, \citet{Winn2018} found that the radius distribution of USP planets shows a decline at a physical radius of 2 $R_{\oplus}$, indicating that USP planets tend to be small (despite discovery biases against smaller planets). Although only a limited number of USP planets have measurements of both their mass and radius, since they tend to be discovered via the transit method, densities are largely consistent with rocky compositions \citep[e.g.,][]{Grunblatt2015, Weiss2016, Vanderburg2017, Christiansen2017, Sinukoff2017, Weiss2021}, and often reside in multi-planet systems \citep{Winn2018, Adams2020}. 

For systems containing many short-period planets, one explanation for the packed configurations is disk migration, which can move the entire planetary systems inward while largely preserving their high multiplicities. Small planets are expected to migrate via Type I migration, which acts through an exchange in angular momentum between a planet and the disk over relatively short timescales \citep{Goldreich1979, Ward1997} and is a rapid process \citep{Bitsch2015, Ogihara2015}. Since small planets are expected to form quickly \citep{Johansen2017}, Type I torques have time to change their orbital locations. 

Given the timescales for both the formation and migration of small planets, the placement of planets near the inner disk edge occurs naturally in the emerging standard picture of planetary system formation.\footnote{In fact, some models of planetary systems indicate that might be harder to prevent planets from reaching the inner edge than it is to move them there \citep{Ogihara2015}.} Once a planet migrates to the disk edge, however, Type I torques vanish (or change sign) and migration comes to a halt  \citep{Masset2006}. In particular, smaller planets may be stopped just outside the disk edge \citep{Tsang2011}, whereas larger gap-forming planets can move all the way to the edge itself. 

Typical disk truncation radii during the T Tauri phase are estimated to be $0.05-0.1$ AU. As a result, disk migration acting alone does not readily populate the region interior to $\sim0.03$ AU with planets. Although larger planets might be able to migrate slightly inside the disk edge before becoming trapped \citep{Tsang2011}, moving smaller planets to the orbital locations where USP planets are observed is more difficult. For this reason, many mechanisms involving planet-planet scattering \citep{Petrovich2019} or tidal migration \citep{Lee2017} have been proposed to explain the observed USP planets with $\sim1$ day orbital periods. These mechanisms require the eccentricity of the USP planets to have been excited in the past, and often leave dynamical signatures of that violent past on the inner system. 

Although routinely applied as a heuristic model, the picture of a disk as a static object with a fixed truncation radius is an oversimplification. In reality, the truncation radius of a disk changes as the star evolves, which is generally a gradual process in which the stellar mass, radius, and magnetic field settle to main sequence values. In addition to these slow changes, however, episodic accretion can quickly and drastically alter the structure of the accretion disk. 
For example, studies of the derived stellar and disk parameters during FU Ori outbursts reveal not only significant brightening \citep{Herbig1966, Herbig1977}, but also accretion rates many orders of magnitude higher than for typical T Tauri stars \citep{Hartmann1996}, altered temperature profiles \citep{Popham1996}, an optically thick structure in a broader waveband \citep{Tapia2017}, and changes in the gas velocity distributions \citep{Shu2007}. All of these variations substantially change the structure of the inner disk region. 

It is well established that T Tauri stars have significant magnetic fields \citep{JohnsKrull2007, Johnstone2014} that thread through the interior regions of the protoplanetary disk. In addition, the inner disk is also influenced by remnant magnetic fields dragged in during protostellar collapse \citep{Lizano2015}, resulting in an extra source of magnetic support that varies with the radial location within the disk \citep{Lubow1994, Shu2007} {and exceeds the strength of the stellar dipole field near the inner disk edge \citep{Ferreira2006}. }
Observations of the mass to flux ratios in maser rings indicate that disks preserve a significant fraction of the magnetic fields of their parent cloud \citep[e.g.,][]{Edris2007}, and once infall has ceased the mass-to-flux ratio will remain fixed. 
{The remnant field dragged in during star formation can persist as planets form and can provide additional pressure support to the disk. The field can also drive physical processes such as accretion, and alter planet formation pathways \citep{Lizano2010, Bai2013, Simon2013}. The specific strengths and longevities of these fields depend sensitively on simulation set-up and parameters \citep[e.g.,][]{Joos2012, Guilet2014, Masson2016}. In this work, we use the 1.5D mean field magnetohydrodynamics and resultant field strengths from \citet{Shu2007}, but it important to note that different assumptions here will alter the results, as weaker disk magnetization will result in more nearly Keplerian gas flow.}

The magnetic fields present in the inner disk {(due to the star as well as the magnetized disk itself}) thus provide a departure from Keplerian rotation for the gas disk, where the effect is largest regions where the magnetic field is strongest \citep{Shu2007, Cai2008}. This magnetic support provides an additional torque on planets orbiting at Keplerian speeds in the sub-Keplerian gas flow \citep{Adams2009}.

Previous work showed that the torque provided by sub-Keplerian rotation profiles can dominate over the Type I torque for short-period planets and lead to faster migration rates, and the effect is particularly relevant in the case of FU Ori outbursts \citep{Adams2009}. Building on these previous results, this paper shows that USP planets can reach their final orbital locations with disk migration alone, which requires accounting for magnetic fields and migration driven by the sub-Keplerian gas flow. In Section \ref{sec:sec2}, we discuss the disk properties for T Tauri and FU Ori systems. In Section \ref{sec:sec3}, we compute the migration forces affecting planets near the inner disk edge, and study the corresponding orbital migration. We focus on planets starting at the inner edge of typical T Tauri disks, with migration driven by sub-keplerian headwinds during FU Ori outbursts, and determine how the results depend on system parameters. In Section \ref{sec:discuss}, we discuss the observational consequences and uncertainties regarding this mechanism. The paper concludes in Section \ref{sec:conclude} with a short summary of our results and a discussion of their implications.

\section{Disk Properties}
\label{sec:sec2}

In order to explain disk-driven orbital migration of exoplanets, it is necessary to construct a working model of the disk where they form and migrate. These disks are generally described in terms of their surface density and temperature profiles. Following astrophysical convention, we we use a power-law surface density profile of the form 
\begin{equation}
\Sigma(r) = \Sigma_{0} (r_{0} / r)^{p},
\label{eq:surfacedensity}
\end{equation}
where $r$ is the radial coordinate of the disk, $\Sigma_{0}$ is the surface density at distance $r_{0}$, and $p$ the power law index. We define the temperature profile of the disk as
\begin{equation}
T(r) = T_{0} (r_{0} / r)^{q},
\label{eq:tempdistribution}
\end{equation}
where $T_{0}$ the disk temperature at a distance $r_{0}$, and $q$ the power law index of the temperature profile. 

The disk scale height depends on the radius of the disk and can be derived from from Equation (\ref{eq:tempdistribution}) to take the form
(e.g, \citealt{Adams2009b}):
\begin{equation}
\frac{H}{r} = \left( \frac{H}{r} \right)_{0} \left( \frac{r}{r_0} \right)^{(1/2 - q/2)}.
\label{eq:hrratio}
\end{equation}
{Here, the benchmark value $(H/r)_{0} = 0.05$ at a distance of $r_{0}$ = 1 AU. As a result, near the inner disk edge at 0.1 AU, we expect the ratio $H/r\sim0.035$ (for $q$ = 3/4).} 

The location of the inner edge of the disk is determined by the balance between disk and stellar properties. More specifically, the disk is truncated at the radial location where the inward pressure due to the disk accretion flow is balanced by the outward pressure due to stellar magnetic fields. Interior to this radius, disk material is cleared out so that a substantial gap exists.\footnote{Although the surface density of this inner region is low, it remains nonzero.} The expression for the truncation radius has been derived previously (\citealt{Ghosh1979,bp1982}; see also \citealt{Ostriker1995,Bouvier2007,Cai2008}) and can be written in SI units in the form 
\begin{equation}
    R_{X} = \Phi_{DX}^{-4/7} \left( \frac{\mu_{*}^4}{G M_{*} \dot{M}^{2}} \right)^{1/7}
    \label{eq:truncationradius}
\end{equation}
where $\mu_{*} = B_{*} R_{*}^{3} / \sqrt{2 \mu_0}$ is the magnetic dipole moment of the star, $\mu_0$ is the vacuum permeability constant, $B_{*}$ is the stellar magnetic field in Tesla, $R_{*}$ the stellar radius, $M_{*}$ is the stellar mass, and $\dot{M}$ is the mass accretion rate onto the central star. The dimensionless scaling parameter $\Phi_{DX}$ is defined as the stellar dipole flux through the disk exterior to the X-point for unpinched fields, where $\Phi_{DX}\sim1$ for reasonable assumptions. In this work, we use the specific value $\Phi_{DX}^{-4/7} = 0.923$, as given in \citet{Ostriker1995}.

We can use typical values for T Tauri stars to estimate the expected truncation radius. For a T Tauri star with a 1 kG magnetic field, a radius of $2.5\ R_{\odot}$, a mass of $1.0\ M_{\odot}$, and a mass accretion rate of $10^{-8}\ M_{\odot}\ /$ yr \citep[see for example][]{Gullbring1998, Herczeg2008}, the truncation radius will be $R_{X} \approx 0.1$ AU.
\citet{Johnstone2014} calculates truncation radii for a selection of stars and finds the range $R_{X} = 2 - 18\ R_{*}$, where the value depends on both the specific T Tauri star and the assumptions made in the magnetic modeling. Earlier work with different assumptions found similar values \citep{Kenyon1996}. 

From Equation (\ref{eq:truncationradius}), it is clear that small changes in stellar and disk properties do not lead to large changes in the truncation radius. Only parameter changes by orders of magnitude will significantly affect the truncation radius. While the disk is present, during the first 1 -- 10 Myr of the star's life, the stellar magnetic field, radius, and mass will not change by large amounts. 
The accretion rate, however, can vary by many orders of magnitude on rapid time scales for a single star. Such variations occur if the star undergoes episodic accretion, in which case the truncation radius will be altered significantly on short timescales. FU Ori outbursts are one type of episodic accretion that are observed around stars with masses between 0.3 -- 1.5 $M_{\odot}$ \citep{Hartmann1996}. 
For nominal T Tauri stars, accretion rates are generally around $10^{-8}\ M_{\odot}$/yr \citep{Herczeg2008, Ingleby2013}, but during FU Ori events the mass accretion rate can increase to $10^{-4}\ M_{\odot}$/yr. As a result, 
the vast majority of the stellar mass can be transferred during short bursts of accretion, and the accretion rate is lower for most of the star's history. 

\begin{figure}
 \includegraphics[width=3.3in]{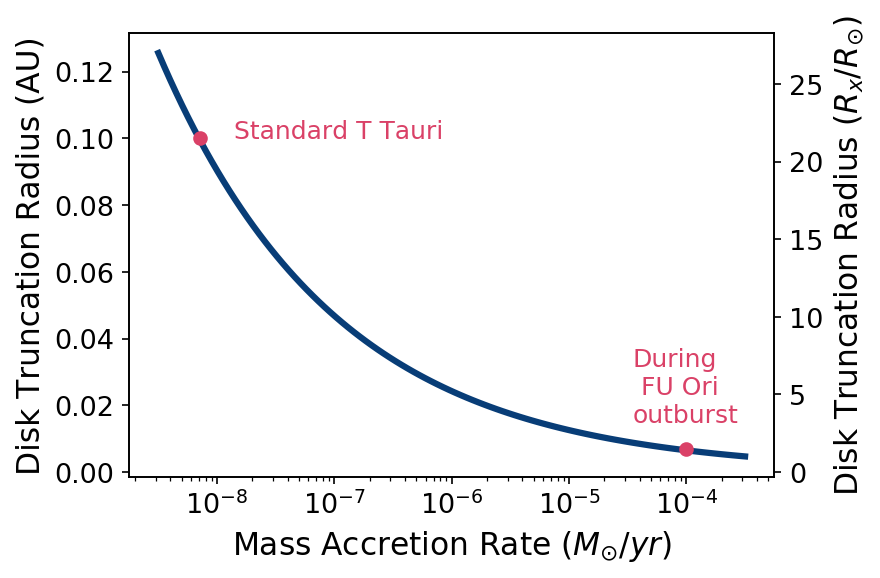}
 \caption{The truncation radius for a disk around a star with magnetic field strength $B_{*} = 1$ kG, radius $R_{*} = 2.5\ R_{\odot}$, mass $M_{*} = 1.0\ M_{\odot}$, and varying mass accretion rates. For larger mass accretion rates, corresponding to an FU Ori outburst event, the disk truncation radius moves inward, potentially near the stellar surface. For more typical quiescent accretion rates, the disk truncation radius will reside around 0.05 -- 0.1 AU. }
    \label{fig:truncation_radius}
\end{figure}

In Figure \ref{fig:truncation_radius}, we plot Equation (\ref{eq:truncationradius}) for a variety of mass accretion rates, with the quiescent and FU Ori outburst rates marked with annotations. In this figure, the stellar parameters are fixed to $B_{*} = 1$ kG, $R_{*} = 2.5\ R_{\odot}$, and $M_{*} = 1.0\ M_{\odot}$.
A change in the mass accretion rate by four orders of magnitude can lead to the disk truncation radius moving inward by a factor of $\sim10$. 
The episodic nature of FU Ori accretion bursts is characterized by short timescales \citep[around 100 years;][]{Hartmann1996} and significant brightening of the star. 
During a single outburst, roughly $10^{-2} M_{\odot}$ of material is accreted onto the star. A star is likely to have multiple outbursts throughout its lifetime, with time between outbursts ranging between $10^4$ - 10$^{5}$ years. 

Although many stars observed to be in FU Ori outburst states are particularly young, some FU Ori stars are classical T Tauri stars, indicating that FU Ori outbursts can occur even after a stellar age of 1 Myr \citep{Miller2011, Hillenbrand2015}. As the same time, planet formation can be a rapid process \citep{Lambrechts2012, Johansen2017}, so that FU Ori events can occur after planet formation has begun or even concluded. As a result, planets at small orbital radii can be strongly affected by the altered disk properties during the outbursts. 
In particular, these episodes provide an opportunity for planets to migrate inward, well past the standard truncation radius. 

\section{Migration Forces Affecting Ultra-Short-Period Planets}
\label{sec:sec3}
Before the protoplanetary disk dissipates, and while the outbursts described above continue to occur, planets are subjected to torques due to interactions between the planet and material in the disk. These torques can alter the orbital elements of the planet, including causing the planet to migrate inwards or outwards (depending on the disk properties). For planets not massive enough to open a gap in the disk, two main sources of torque lead to migration in the inner disk: Type I torques and the torque provided by a headwind driven by sub-Keplerian gas velocities. The latter are relevant everywhere that magnetic fields provide significant support for the rotation curve of the gas. However, the relative importance of the two torque contributions depends on semimajor axis, where the headwind torque dominates in the innermost regions \citep{Adams2009}.

\subsection{Type I Torques}

The magnitude of the torque exerted on the planet due to net Lindblad and co-rotation torques \citep{Goldreich1979, Ward1997} has the form 
\begin{equation}
    \Gamma_{I} = - C_{I} \left( \frac{m_p}{M_{*}}\right)^2 \pi \sigma a^{4} \Omega_{kep}^{2} \left( \frac{a}{H}\right)^2\,,
   \label{eq:torque_typei}
\end{equation}
where $a$ is the semi-major axis of the planet, $\Omega_{kep}$ is the Keplerian rotation rate at orbital radius $a$, and $H$ is the disk height at radius $a$ (the ratio $a/H$ varies with orbital radius and is set by Equation [\ref{eq:hrratio}]). The dimensionless coefficient $C_{I}$  depends on the disk properties and determines the direction of the torque. This torque leads to inward or outward motion of the planet \citep{Goldreich1979}, depending on disk properties as parameterized by the coefficient $C_{I}$ \citep[ex:][]{Paardekooper2010, Bitsch2017}. If the torque is positive, which occurs in many scenarios \citep[e.g.,][]{Dittkrist2014}, including when radiative cooling is not efficient \citep{Paardekooper2006, Yamada2012}, the planet will migrate outwards. 
If the torque is negative, then the planet migrates inward. Numerical experiments \citep{Bitsch2014VIII} indicate that that for planets with smaller masses ($m_p < 10 \ M_{\oplus}$), Type I migration will generally be inward. For many cases, $C_{I}$ can be computed using only the disk profile exponents $p$ and $q$ \citep{Tanaka2002, DAngelo2010, Paardekooper2010}. For locations in the disk where $p=3/2$, $q = 3/4$, Type I migration will tend to move small planets inwards \citep{Bitsch2014VII, Bitsch2014VIII}.

However, the typical scenario where Type I torques move planets inward  does not hold once planets reach regions slightly exterior to a disk cavity (including the central cavity inside the truncation radius).  Numerical simulations \citep{Masset2006} indicate that near a cavity edge, where the disk surface density profile experiences a steep gradient, increasingly positive corotation torques will cancel out the negative differential Lindblad torques and provide the planet a positive net torque. The radius at which this occurs is often called a planet trap, since the torque balance will cause a planet to halt migration and become trapped at a fixed orbital radius. This behavior has been found in several other numerical studies \citep[e.g.,][]{Morbidelli2008, Baillie2016, Romanova2019}.
Since the planets we consider in this work reside at (or near) the inner disk edge, slightly external to the interior cavity, Type I torques are expected to act outwards and {under standard assumptions} $C_{I} \approx - 0.6$ \citep{Ward1997, Tanaka2002}. 
{However, as shown in \citet{Paardekooper2009},  sub-keplerian gas flow can alter the relative positions of the Lindblad resonances (see their Figure 2), and $C_{I}$ must be adjusted according to the disk scale height and the velocity of the gas flow. In general, this modification leads to a stronger Type I torque than predicted by standard models, with the exact scaling determined by the local disk surface density profile and disk aspect ratio. Nonetheless, in the inner disk regions, the Type I torque remains subdominant (see below). For the sake of definiteness, this work assumes $C_{I} \approx - 1$, but keep in mind that this value may vary for different disk parameters.  }

\subsection{Torque due to Sub-Keplerian Gas Velocities}
As discussed in Section \ref{sec:sec2}, magnetic fields thread through the protoplanetary disk and provide pressure support. The magnetic fields originate from two sources: a stellar magnetic field that scales with radius as $B(r) \propto  r^{-3}$ and the remnant field dragged in during disk formation, which scales as $B(r) \propto r^{-5/4}$ \citep{Shu2007}. These fields cause the gas to orbit at a sub-Keplerian rate. The torque that a sub-Keplerian gas disk exerts on a planet \citep{Adams2009} can be written in the form 
\begin{equation}
    \Gamma_{X}  = - C_{D} \pi r_{p}^2 a \rho_{gas} v_{rel}^{2}\,,
\end{equation}
where the planet itself is assumed to move at a Keplerian velocity. In this expression, $C_{D}$ is a dimensionless drag coefficient analogous to $C_{I}$, but has a value $C_{D} \sim 1$ for a planet in a disk.\footnote{More specifically, we note that the drag coefficient is a slowly varying function of Reynolds number, but remains of order unity up to large values (e.g., \citealt{fluidbook}).} As before, $r_{p}$ is the planetary radius, $\rho_{gas} = \Sigma / 2H$ is the gas density, and $v_{rel}$ is the velocity differential between the gas and planet velocities, defined as $v_{rel} = v_{kep} - v_{gas}$. Following \citet{Shu2007}, we define the parameter $f$ to be the ratio between the true rotation rate of the gas and the expected Keplerian rotation rate, $f \equiv \Omega_{gas} / \Omega_{kep}$. For planet in a a Keplerian orbit in a disk rotating at a Keplerian speed, $f = 1$. 
This parameterization allows us to rewrite $v_{rel} = (1 - f) v_{kep}$. With these substitutions, the torque due to the headwind of a sub-Keplerian gas flow takes the form 
\begin{equation}
    \Gamma_{X} = - C_{D} \pi r_{p}^2 (\Sigma / 2H) (1 - f)^{2} G M_{*} .
    \label{eq:torque_xwind}
\end{equation}

\subsection{Total Torque on a Planet}
The relative importance of the two contributions to the total torque (Equations  [\ref{eq:torque_typei}] and  [\ref{eq:torque_xwind}]) depends on the parameters of the planet. To compare the importance of these contributions for a particular object, we must assume a mass-radius relation for the planets. We construct a piece-wise continuous mass-radius relation by computing the planetary bulk density $\rho_p$ for three regimes: 
\begin{equation}
\begin{aligned}
\rho_p = 
\begin{cases}
2.43 + 3.39\ r_{p}\ \rm{g / cm}^3 \ \ \ {\rm if}\ r_{p} < 1.5 R_{\oplus} \\
14.84\ r_{p}^{-1.7}\ \rm{g / cm}^3\ \ \ {\rm if}\ 4 > r_{p} > 1.5 R_{\oplus} \\
1.45\ \rm{g / cm}^3\ \ \ {\rm if}\ r_{p} > 4 R_{\oplus} \\
\end{cases}
\end{aligned}
\label{eq:mass-radius}
\end{equation}
The expression for planets with $r_{p} <1.5 R_{\oplus}$ is taken from the \citet{Weiss2014} mass-radius relation, the expression for planets with $4 R_{\oplus} > R_{p} > 1.5 R_{\oplus}$ is derived from the \citet{Wolfgang2015} relation. For both cases, we ignore the possible range due to uncertainties and use only the best-fit line from each paper to simplify the calculation. For planets larger then $4 R_{\oplus}$ the density is taken to be a constant $\rho = 1.45$ g / cm$^{3}$, which matches the upper end of the previous range. Using this conversion, combined with values of 
$C_{I} = -0.6$ for a T Tauri star \citep{Shu2007}, 
$C_{D} = 1$ \citep{Weidenschilling1977b}, $M_{*} = 1 M_{\odot}$, for a disk with $p=3/2$, $q = 3/4$, we can compute the ratio of torques for a variety of planet masses and orbital separations. 

\begin{figure}
 \includegraphics[width=3.3in]{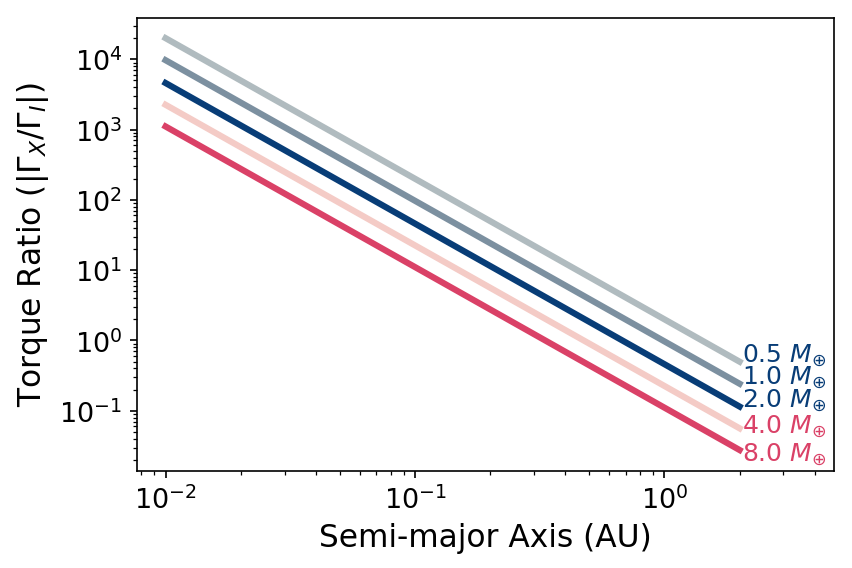}
 \caption{For a selection of planetary masses, we plot the ratio between the size of the torques provided by the sub-Keplerian gas disk when $f = 0.658$ and by the Lindblad and corotation torques over a range of semi-major axis. For low mass planets and those at shorter orbital separations, the sub-Keplerian gas disk provides a larger contribution to the torques acting on the planet. Beyond 1 AU, the sub-Keplerian contribution becomes negligible compared to the Type I torque. }
    \label{fig:torque_ratio}
\end{figure}

The resulting ratio between the magnitudes of the Type I and headwind-driven torques is shown in Figure \ref{fig:torque_ratio} for a range of orbital locations and planet masses. 
The torques arising from the sub-Keplerian gas disk are important for close-in planets and those with lower planetary masses. For planets at moderate orbital separations, $\sim1$ AU, the torque reduces to the Type I equation for nearly all planet masses. 

Figure \ref{fig:torque_ratio} shows that even though torques due to the headwind from sub-Keplerian gas flow are not important beyond 1 AU, they can be important for planets at short orbital separations. For small planets at 0.05 AU and within, the magnitude of the Type I torque can be several orders of magnitude smaller than that provided by the sub-Keplerian gas disk.

\subsection{Migration Timing}
For planets to be affected by the headwind-driven torque, they must reside at fairly small orbital radii. To determine how often such torques are operative, we briefly describe the relative time scales. 
{Planetesimals can form more rapidly than the grain migration timescale \citep{Johansen2007} through the streaming instability \citep{Youdin2005, Squire2018}, while a magnetized disk may accelerate this process \citep{Seligman2019}.}
{Subsequently, planet formation from planetesimals} can take place rapidly, but it is unlikely that planets will form exactly at the inner edge of the disk \citep{Raymond2020}. Planets residing near the disk edge are more likely to have formed further out in the disk, and then migrated inward. As a result, the limiting timescale for placing planets at the disk edge will be their migration time. Here, as before, we consider smaller planets that migrate via Type I migration. 

The timescale of Type I migration can be estimated using the result of \citet{Tanaka2002}, 
\begin{equation}
    t_{I} = \frac{1}{4.35}\  \frac{M_{*}^2}{\Omega\ m_p\ \Sigma\ a^2} \left(\frac{H}{a}\right)^2\,.
\end{equation}
The numerical coefficient is computed assuming a surface density profile with $p = 3/2$. For an Earth-mass planet that forms at $a=1-2$ AU, this timescale is of order $6-8 \times 10^4$ years for a surface density of 2000 g cm$^{-3}$. Type I migration is rapid compared to disk dissipation time scales \citep{jesus}. Provided that a planet forms early, Type I torques will generally be able to bring it close to the inner edge of the disk.

\begin{figure}
 \includegraphics[width=3.4in]{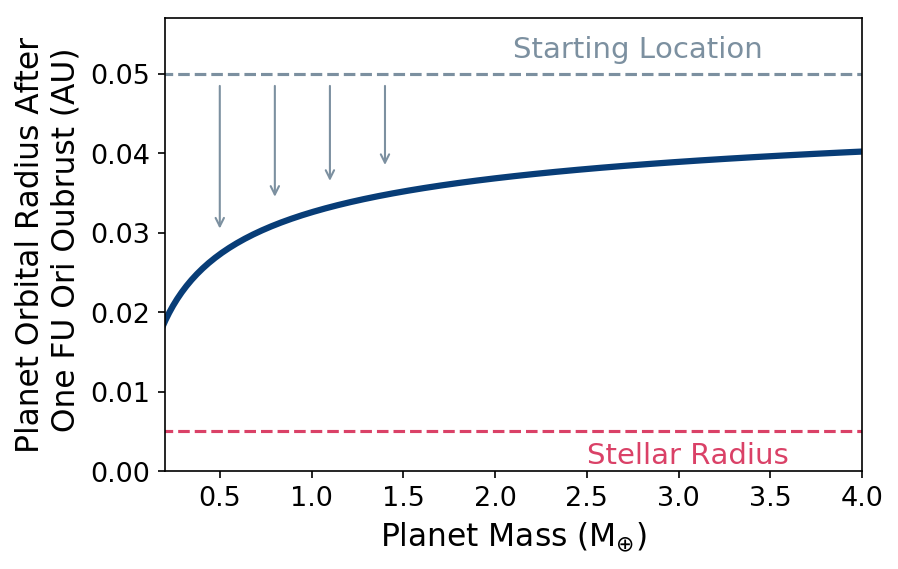}
 \caption{For planets that reside at initial orbital distance 0.05 AU during the standard T Tauri accretion phase, this figure shows the final orbital distance that can be reached after one FU Ori outburst while $f = 0.386$ \citep{Shu2007}. The final state radius is plotted versus the planet mass. For planets with smaller masses, the headwind provided by the sub-Keplerian rotation curve acts to rapidly migrate planets inward. The orbital radius can decrease by a factor of $\sim$2.5 during a single FU Ori event.}
    \label{fig:migration_distance}
\end{figure}

Although they move the disk edge inward, FU Ori outbursts are several orders of magnitude too short for a planet to migrate from 0.05 AU to an ultra-short-period orbit at 0.01 or 0.02 AU via Type I migration. More importantly, because the inner disk edge has a repulsive effect to small planets, Type I torques alone cannot force a planet to remain at 0.01 AU. 
From Figure \ref{fig:torque_ratio}, however, it is clear that at short orbital periods, the sub-Keplerian flow-driven torque will be much larger than the Type I torque. In this case, even though Type I migration rates are not fast enough to successfully move a planet down to $\sim0.01$ AU in a few hundred years (over an FU Ori outburst), the combined effect of the two mechanisms might be sufficient. To assess this possibility, we compute the distance that a planet can migrate under the combined torque from both sources using the differential equation 
\begin{equation}
    \frac{da}{dt} = 2 \frac{\Gamma_{t}}{a m_{p}} \sqrt{a^3 / G M_{*}} \,,
\label{eq:dintegral}
\end{equation}
where $\Gamma_{t} = \Gamma_{I} + \Gamma_{X}$ is the total torque acting on the planet, and can be computed using Equations (\ref{eq:torque_typei}) and (\ref{eq:torque_xwind}). For a typical T Tauri accretion disk, the sub-Keplerian factor near the disk edge has been estimated to be
$f =  0.658$, whereas for a star/disk system undergoing an FU Ori outburst, $f =  0.386$ \citep{Shu2007}. 
In this scenario, a planet formed at a moderate radius in the disk (1 -- 2 AU) migrates inward fairly rapidly. It will subsequently reside near the disk inner radius, and during FU Ori outbursts be subject to headwind torques as the disk edge moves further inward according to Figure \ref{fig:truncation_radius}. 

In Figure \ref{fig:migration_distance}, we plot the distance that a planet starting at 0.05 AU would migrate during one FU Ori outburst, computed by solving Equation (\ref{eq:dintegral}) for $t_{max} = 100$ years (corresponding to the lifetime of a single FU Ori outburst) and the standard star/disk parameters that we have used thus far in this work ($f =  0.386$ for a FU Ori outbursting star, ${\dot M} = 5 \times 10^{-4}\ {M_\odot}/$ yr, $M_{*} = 1 {M_\odot}$, $p=3/2$, $q = 3/4$, $C_{I} = -0.6$, and $C_{D} = 1$). 
In order for a planet to reach the original inner disk edge, it must form and migrate inward some distance, which requires a good fraction of the disk lifetime. As a result, due to the relative infrequency of FU Ori outburst events, by the time a planet reaches this location, only a few FU Ori outbursts may subsequently occur.
We note that this process may not be broadly applicable due to both the timing of planet formation, migration, and FU Ori events, as well as the particular parameters of individual systems.  
This mechanism may affect only a minority of systems. 

Figure \ref{fig:migration_distance} shows that planets with smaller masses migrate more readily to smaller orbital radii during an FU Ori event. Qualitatively, this makes is sensible; the effect of a headwind provided by a sub-Keplerian gas disk will be more important for smaller planets (unlike Type I torques, which increase in size as planet mass increases). For the smallest planets, those will masses below 1 $M_{\oplus}$, a single FU Ori event could provide sufficient time for a planet to migrate into an ultra-short-period orbit (at $\sim0.02$ AU) while the disk truncation radius is temporarily reduced according to Equation (\ref{eq:truncationradius}) due to the increased mass accretion rate. 

\begin{figure}
 \includegraphics[width=3.3in]{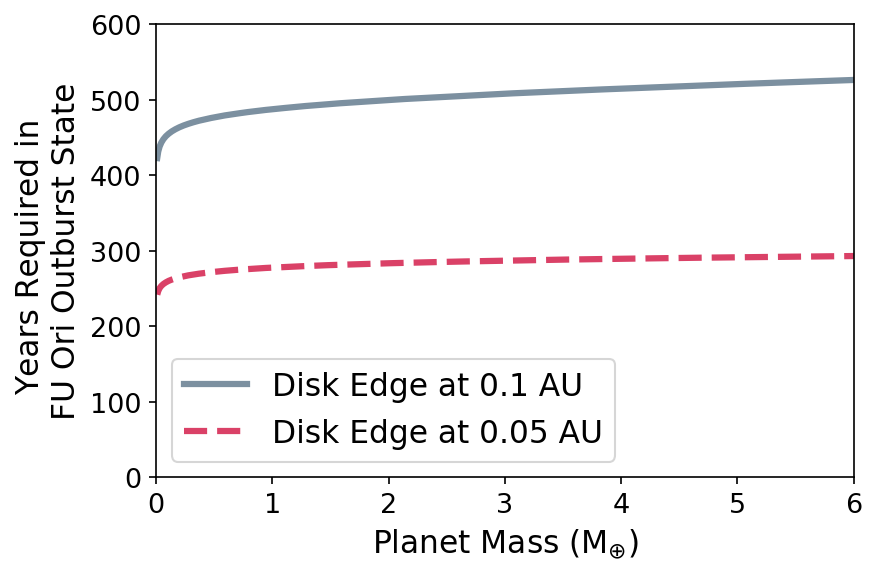}
 \caption{For a planet of varying mass residing at a disk edge of 0.1 AU (top line) or 0.05 AU (bottom line) during the initial quiescent T Tauri phase, we plot the number of years the star must reside in a FU Ori outburst state in order to provide enough time for the planet to migrate inward to final orbital radius 0.01 AU. The smallest planets experience larger migration distances for a fixed time as computed by Equation (\ref{eq:dintegral}), and thus reach smaller final orbits for a given number of FU Ori events. }
    \label{fig:years}
\end{figure}

Considering that there is no canonical number of FU Ori events that a planet may experience, in Figure \ref{fig:years}, we plot the time (in years) that a planet must reside at or within the disk truncation radius and reside in a disk subject to FU Ori-like mass accretion rates in order for the planet to migrate into a USP orbit. These results were again computed using Equation (\ref{eq:dintegral}) and the same stellar and planetary parameters specified previously. We assume that after a single FU Ori event, the inner radius of the disk recedes quickly and decouples from the planet, which is left at its instantaneous orbital radius. When the next FU Ori event starts, the gas disk again rapidly decreases its inner radius (past the planet) and the headwind migration effect continues to migrate the planet further inward (it picks up where it left off). 
This piecemeal migration is possible in the picture of headwind-driven migration but not for Type I migration alone.

Although the specifics will vary, Figure \ref{fig:years} indicates that in order for a planet to reach an  ultra-short-period orbit (0.01 AU), more than one FU Ori outburst is often required. Intuitively, disks with inner edges closer to the star during the T Tauri phase will produce USP planets more readily. This process also has a weak dependence on mass, where larger planets require a longer phase of FU Ori events to reach USP orbits. Through this mass dependence, headwind migration predits a higher abundance of smaller planets at the smallest orbital radii. 

\section{Discussion}
\label{sec:discuss}
The concentration of magnetic fields at small orbital radii during mass accretion leads to significantly sub-Keplerian rotation. 
This paper demonstrates that migration driven by such sub-Keplerian rotation curves in the inner disk allows planets (preferentially smaller planets) to migrate inward during FU Ori events (building on the work of \citealt{Adams2009}). The sub-Keplerian flows themselves are provided by the strong magnetic fields present in the disks of T Tauri stars. Through this mechanism, ultra-short-period planets can be produced by disk migration and a variable disk truncation radius. The disk truncation radius will shrink as the mass accretion rate of the star changes during FU Ori outbursts, allowing planets that would normally be trapped near $\sim0.05$ AU to migrate inward during the short duration of the FU Ori event.
This mechanism invokes no dynamical instability, and requires only that a planet reaches the truncation radius of the disk and the star subsequently experiences at least one FU Ori accretion event. 

\subsection{Specifics on the Gas Flow Velocity} 
\begin{figure}
 \includegraphics[width=3.3in]{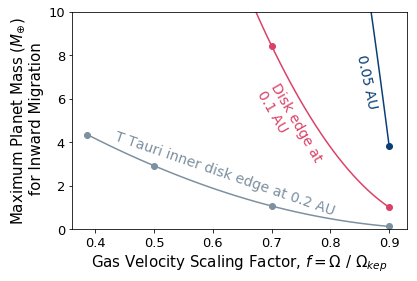}
 \caption{Near the disk edge, Type I torques act outwards whereas headwind-driven torques act inwards. The result of this compromise is that planets only migrate inward if the combination of their physical size and the sub-Keplerian gas velocity allow the headwind torques to overcome Type I migration. The ratio $f$ of the gas velocity to planet velocity determines the maximum planetary mass that can migrate inward during FU Ori outbursts. {This figure shows this maximum planet mass for a disk for starting locations of 0.05, 0.1 and 0.2 AU (the disk edge during the T Tauri phase). }}
    \label{fig:minmass}
\end{figure}

This migration mechanism requires that the gas in the inner disk orbits with a substantial departure from a Keplerian rotation curve. \citet{Shu2007} estimated the amount by which the gas velocity would be sub-Keplerian using typical values for the star and disk parameters in various phases of evolution (T Tauri, FU Ori, and protostars). However, the scaling factor $f$ is sensitive to the amount of magnetic fields dragged in, the disk surface density, and the mass-to-flux ratio in the disk, so that $f$ will vary 
from system to system. A smaller value of $f$ (greater departure from Keplerian rotation) makes the mechanism outlined in this paper more effective. But even if the gas flow is substantially more Keplerian (larger values of $f$), the smallest planets will still be affected.  Figure \ref{fig:minmass} shows the maximum planet mass that can be pushed inward by headwind-driven torques for a variety values of the sub-Keplerian factor $f$. 

\subsection{The Effect of Episodic Accretion}
In this paper, we assumed that two accretion states were possible for the young systems: a normal T Tauri accretion rate of $10^{-8} M_{\odot}/$ yr, and an enhanced accretion rate during FU Ori outbursts, where the accretion rate increases to $\sim 10^{-4} M_{\odot}/$ yr for a short duration ($\sim$100 yr). Each accretion rate corresponds to a different disk truncation radius (see Figure \ref{fig:truncation_radius}). Moreover, if planets can migrate quickly enough during the FU Ori outburst, they can reach the target orbital periods of less than a day to become USP planets. {As shown in Figure \ref{fig:minmass}, if the inner disk edge is closer to the host star during the quiescent phase of mass accretion, the system will form USP planets more readily (and possibly increase their chances of subsequent destruction). In contrast, larger values of the inner disk edge may significantly limit the masses of planets allowed to move inward through this mechanism.}

However, the accretion rate must vary between the two aforementioned extremes as the disk evolves. A variety of processes have been proposed which will increase the accretion rate to some degree, but not to FU Ori outburst levels (e.g., MHD turbulence \citealt{Armitage2001}). In addition, numerical modeling has shown that in systems with FU Ori outbursts, mass accretion rates vary both slowly and stochastically as the star evolves \citep{Vorobyov2006}. These modes of disk operation move the truncation radius inwards slightly, and allow planets a closer starting position compared to the 0.05 AU that we assumed in this work. This head start allows planets to migrate to the 0.01 -- 0.02 AU range through a smaller number of FU Ori outbursts. Additionally, as mentioned in \citet{Adams2009}, another effect of episodic, rapid increases in the accretion rate is that the star cannot spin up fast enough, potentially contributing additional inwards migration.

Additional work \citep[as outlined in][]{Hillenbrand2015} to classify the specific parameters of FU Ori outburst stars, including the structure and temperature of the inner accretion disk and the frequency and duration of their outbursts (Rodriguez et al., \emph{in prep}), will improve estimates of how frequently the mechanism outlined in this work can affect the orbits of short-period planets. Identification of new FU Ori stars \citep{Hillenbrand2018} will also improve our understanding of the demographics of sources that undergo FU Ori outburst events. 

\subsection{Observational Consequences}
\begin{figure}
 \includegraphics[width=3.3in]{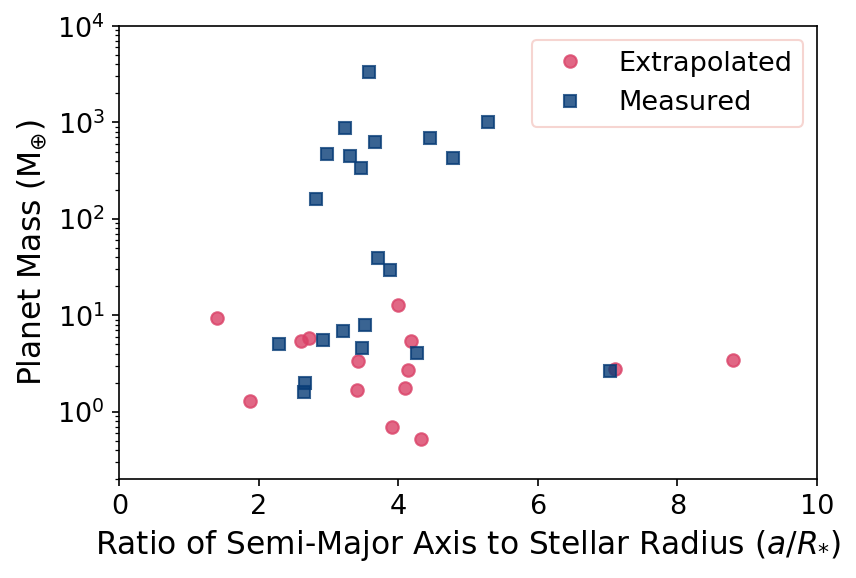}
 \caption{The masses of confirmed ultra-short period planets plotted versus their orbital separation in units of stellar radii. For systems where measured masses were available, true mass is plotted. For systems where mass has not been measured (mostly planets discovered via the transit method), the mass has been extrapolated from the observed radius using Equation (\ref{eq:mass-radius}). Although the number of planets in this sample is low and observational biases tend to prevent the discovery of the smallest planets, the planets at the shortest orbital separations tend to be small. Parameters for planets on this plot come from \citet{Gillon2014, SanchisOjeda2015c, Esteves2015,Wong2016,  Stassun2017,Oberst2017,Vanderburg2017, Livingston2018a, Livingston2018,  Bourrier2018,  Smith2018, Malavolta2018, Rodriguez2018, Hirano2018, 
Mayo2018, Chakrabarty2019, Vanderspek2019, Vines2019, Frustagli2020, Nielsen2020, Cloutier2020, Jenkins2020, McCormac2020, Armstrong2020, Weiss2021, Lacedelli2021}.}
    \label{fig:data}
\end{figure}

The mechanism outlined in this work makes distinct observational predictions regarding the mass distribution of ultra-short-period planets. 
Headwind-driven migration will operate for small planets that are more affected by the velocity differential with the gas disk. Larger planets that migrate via Type II migration and/or clear a gap will not be affected by this mechanism. Figure \ref{fig:data} shows the observed distribution of orbital separation and planet mass for confirmed planets with orbital periods less than 1 day. Although many observational biases have been folded into this diagram (in particular, smaller planets are more difficult to find, so that completeness is lower for lower mass planets), the planets with the smallest $a/R_{*}$ ratio do appear to be those with smaller masses. As the plane shown in Figure \ref{fig:data} becomes populated with a greater number of exoplanets, and detected planets have improved observational characterization, more quantitative tests of the mechanism in this paper will be possible.

The underlying physical parameters that define this migration scenario are subject to uncertainties. This work shows that magnetically-enforced sub-Keplerian rotation curves will create some mass dependence for the occurrence rates of USP planets around stars that host FU Ori outbursts. However, the inner radius to which a planet can migrate depends sensitively on the properties of the system, including the magnetic field strength of the host star, the mass and radius of the planet, the surface density of the disk, and the duration and mass accretion rates of the FU Ori outbursts. As a result, a more complete assessment of the duty cycle of this migration mechanism requires more accurate determinations of those parameters. 

As one example, consider a 1 $M_{\oplus}$ planet starting at 0.05 AU, and typical values for the range of star, disk, and planet parameters. During a single FU Ori event, such a planet could migrate inward to anywhere between 0.02 AU and 0.04 AU. The smaller final orbital radius results from disks with higher overall surface densities, higher $H/r$ ratios, and for stars with higher masses. Since these parameters vary widely over the population of T Tauri stars and their planets, the mechanism discussed in this work will operate to varying degrees in different systems. 

Another complication arises for planets that reach the disk truncation radius too early. During sufficiently powerful FU Ori events, the disk truncation radius can reach the stellar surface. Any planet that experiences too many FU Ori events while residing on a short-period orbit could migrate inwards too rapidly and collide with the star. It is likely that a sizeable fraction of planets migrating via the mechanism described in this paper could meet this end. If planetary engulfment occurs too often, however, the planets would provide a source of heavy elements to the stellar photosphere, leading to metallicity enhancements that are potentially observable \citep{Laughlin1997}. Since observations show that planetary architectures may correlate with stellar metallicity \citep{Brewer2018, Anderson2021}, pinning down the details of engulfment will serve as an important constraint on these processes. 

Tidal dissipation provides another mechanism that can remove USP planets after they form and migrate to their final orbital positions. Tidal interactions with the central star act to move planets inwards after FU Ori outbursts have finished and the mechanism of this paper is no longer operative. 
{Energy could be tidally dissipated within the USP planet itself or in its host star. 
The larger radius of the host star at early times increases the inward migration rate, as the tidal migration rate scales with $Q_{*} R_{*}^5$ \citep{Goldreich1966}. This scaling may lead to the potential destruction of USP planets over time as their orbits decay. }

Another observational diagnostic for USP formation mechanisms is the USP planet occurrence rate as a function of stellar age. In the scenario presented in this paper, USP planets attain their final orbital positions early, while the protoplanetary disk is still present. Although tidal effects \citep{Goldreich1966, Rasio1996} could subsequently remove some planets with appropriate physical properties (see above), the smaller planets more likely to migrate to USP orbits via our proposed mechanism should be found at approximately equal rates around young stars and old stars \citep[consistent with the result of][which found no evidence for statistically different ages for USP planet-hosts and field stars]{Hamer2020}.
In multi-planet systems, migration driven by obliquity tides can take the innermost planet in the system to an USP orbit \citep{Millholland2020}. 
Several other theories also involve dynamical interactions between the USP planet and its companions, including high-eccentricity migration \citep{Petrovich2019} and low-eccentricity migration \citep{Pu2019}. 
All of these mechanisms predict a stellar-age dependence of USP planetary occurrence, and barring eccentricity tides the other mechanisms also require the USP planet to reside in a multi-planet system. 
In contrast, the mechanism described in this work allows for the presence of nearby planetary companions but does not require them, and will not generate a stellar-age dependence on the USP planet occurrence rate.

Finally we note that a number of low probability events that must occur for this mechanism to operate. The planet must form early and migrate to the inner disk edge in time to experience at least one FU Ori event. The magnetic field structure must result in sufficient sub-Keplerian rotation. And the resulting migration mechanism cannot move the planet too far inward --- otherwise the planet will not survive. As a result, we do not expect a large population of planets to be placed on ultra-short-period orbits through this mechanism. USP planets should thus be rare, a finding that is consistent with observational estimates of the USP planet occurrence rate, which is only $\sim 0.5\%$ \citep{SanchisOjeda2014}.

\section{Conclusion}
\label{sec:conclude}

This paper presents a new mechanism to place planets on ultra-short-period (USP) orbits. The key issue is that observed USP planets reside on orbits that lie inside the inferred inner truncation radii of circumstellar disks. This mechanism resolves this issue through the action of two physical processes: FU Ori outbursts can temporarily move the inner edge of the disk inward, thereby allowing planets to migrate in principle. In practice, standard Type I migration rates are too slow, given the limited duration of the outburst phase. If the disk has sufficiently strong magnetic support, the rotation curve can be sub-Keplerian, and the resulting headwind can drive planetary migration in the (short) allotted span of time. 

As delineated in this paper, planets can be moved to USP orbits for a reasonable range of system parameters (see Figures \ref{fig:migration_distance} and \ref{fig:years}). The disk truncation radius naturally decreases from $\sim0.1$ AU (for T Tauri stars) down to $\sim0.01$ AU during FU Ori outbursts (Figure \ref{fig:truncation_radius}). These values require magnetic field strength $B_\ast\sim1$ kG, quiescent ${\dot M}_d\sim10^{-8}$ $M_\odot$/yr, and FU Ori mass accretion rates of order $10^{-4}M_\odot$/yr, where all of these values are observed. 

The other key feature of this mechanism is that the disk must have a sub-Kelerian rotation curve, which requires magnetic fields to be embedded within the disk (in addition to the fields originating from the stellar surface). Such fields and their consequences are expected theoretically, as magnetic field lines are dragged into the disk during its formation \citep{Shu2007}, but direct observational confirmation remains elusive. As the disk rotation curve becomes more Keplerian ($f$ closer to unity), the effects of headwind migration are compromised and fewer planets can move into USP orbits (Figure \ref{fig:minmass}). 
{The exact masses of planets that will be able to migrate inwards depends on the strength of Type I torques, as stronger torques will decrease the minimum mass of planet allowed to migrate inwards for a given gas velocity profile.}

Smaller planets are more readily moved through headwind migration. As a result, the collection of USP planets is predicted to have smaller masses than the overall planetary population. This finding is consistent with currently available data (Figure \ref{fig:data}), but more observations are needed. Even for small planets, successful placement into USP orbits requires a number of specific conditions (e.g., rapid formation and migration to the initial disk edge, with the right amount of remaining FU Ori activity). As result, we expect USP planets to be rare.


%
\medskip
\textbf{Acknowledgements.}  
J.C.B.~has been supported by the Heising-Simons \textit{51 Pegasi b} postdoctoral fellowship.
We thank Tony Rodriguez for useful conversations.
We also thank the anonymous referee for useful comments. 
This research has made use of NASA’s Astrophysics Data System.

Software: pandas \citep{ mckinney-proc-scipy-2010}, matplotlib \citep{Hunter:2007}, numpy \citep{oliphant-2006-guide}, Jupyter \citep{Kluyver:2016aa}

\end{document}